\documentclass[12pt,a4paper]{article}

\usepackage{amsfonts,amssymb,amsmath,amsopn,amsthm,graphicx}

\title{Extended Edge States in Finite Hall Systems}
\author{Christian Ferrari and Nicolas Macris}
\date{Institute for Theoretical Physics \\ Ecole Polytechnique F\'ed\'erale \\
CH - 1015 Lausanne, Switzerland}

 \numberwithin{equation}{section}

\newtheorem{thm}{Theorem}

\newtheorem{lem}{Lemma}

\newtheorem{prop}{Proposition}

\newtheorem{hyp}{Hypothesis}

\newcommand{\D}{\,\textrm{d}}
\newcommand{\Le}{\left}
\newcommand{\Ri}{\right}
\newcommand{\bs}{\boldsymbol}
\newcommand{\N}{\mathbb{N}}
\newcommand{\Z}{\mathbb{Z}}

\newcommand{\R}{\mathbb{R}}
\newcommand{\C}{\mathbb{C}}

\DeclareMathOperator{\supp}{supp}
\DeclareMathOperator{\dist}{dist}

\DeclareMathOperator{\Tr}{Tr}
\begin{document}

\maketitle

\abstract{We study edge states of a random Schr\" odinger operator
for an electron submitted to a magnetic field in a finite
macroscopic two dimensional system of linear dimensions equal to
$L$. The $y$ direction is $L$-periodic and in the $x$ direction
the electron is confined by two smoothly increasing parallel
boundary potentials. We prove that, with large probability, for an
energy range in the first spectral gap of the bulk Hamiltonian, the
spectrum of the full Hamiltonian consists only on two sets of eigenenergies
whose eigenfuntions have
average velocities which are strictly
positive/negative, uniformly with respect to the size of the system. Our result gives a well defined
meaning to the notion of edge states for a finite cylinder with two boundaries, and 
 extends previous studies on  systems
with only one boundary.}


\newpage

\section{Introduction}

In this paper we investigate spectral properties of random Hamiltonians
describing the dynamics of a
spinless quantum particle on a cylinder of circumference $L$
and confined along the cylinder axis by two boundaries
separated by the distance $L$. The particle is subject to 
an external homogeneous magnetic field and a weak
random potential. A
precise statement of the model is given in section 2. The physical
interest of the model comes from the integral quantum Hall effect
occurring in disordered two dimensional electronic systems subject to
a uniform magnetic field, for example, in the
interface of an heterojunction \cite{vKDP}, \cite{PG}. In his treatment
of this effect Halperin \cite{H} pointed out the fundamental role
 played by edge states carrying boundary diamagnetic
currents, and it is therefore important to understand the
spectral properties of finite but macroscopic quantum Hall samples with
boundaries. A short review of the spectral properties of finite
quantum Hall systems can be found in \cite{FM2}.\\

The study of random magnetic Hamiltonians with boundaries is
recent and, before we adress the case of a (finite) cylinder, we wish to briefly discuss a few existing results.
The case of a semi-infinte plane with one planar boundary, modeled
by a smooth confining potential $U$ or a Dirichlet condition at $x=0$, is  satisfactorily understood.
In this case it is proven that the spectrum of the
Hamiltonian $H_\omega^e=H_L+U+V_\omega$, $H_L$ being the Landau Hamiltonian for
a uniform magnetic field $B$ and $V_\omega$ an Anderson-type random potential, has absolutely continuous
components inside the complement of Landau bands, for
$\|V_\omega\|_\infty\ll B$ 
(\cite{FGW}, \cite{dBP} and \cite{MMP}). The proof of this statement is
essentially based on Mourre theory with conjugate operator $y$.
The positivity of $i[H_\omega^e,y]$ in suitable spectral
subspaces of $H_\omega^e$ leads to the absolutely continuous nature
of the spectrum. Since this commutator is equal to the velocity $v_y$ this means that states
in the corresponding spectral subspaces propagate in the $y-$direction along
the edge with positive velocity.

For the case of a strip with two boundaries, separated by a distance $L$,
few results are known.
For a general (random) potential we expect that there is no absolutely continuous component in
the spectrum, because the impurities may induce a tunnelling (or backscattering) between the
two boundaries and thus propagating edge states along each
boundary cannot persist for an infinite time. In \cite{CHS} the
authors have shown that such states survive, for a finite
time related to the quantum tunnelling time between the two edges.
In \cite{EJK} instead of a strip of size $L$, the authors consider 
a parabolic channel. They show  that if the perturbation $V$ is periodic, or
if $V$ is small enough and decays fast enough in the $y-$direction,
then the absolutely continuous spectrum
survives in certain intervals, but their analysis does not cover true Anderson like potentials.\\

In this work we address the case of a macroscopic
finite systems with two confining walls separated by a distance $L$ along
the $x-$direction and with the $y-$direction of length $L$ made periodic 
(i.e. the geometry is that of a cylinder). 
The \emph{left} (resp. \emph{right}) \emph{walls} are modeled by a smooth confining potential $U_\ell$
(resp. $U_r$) sepatated by a distance $L$, and the \emph{bulk} between them contains impurities modeled by a
random Anderson-like potential $V_\omega$.
In
this case, although the spectrum consist of discrete isolated eigenvalues , 
we show that there is a well defined notion of edge states
associated to each boundary.

Let us explain our main result expressed in Theorem \ref{thm1}. We show that, with large probability, the
spectrum of the random Hamiltonian 
$$
H_\omega=H_L+V_\omega +U_\ell+U_r
$$
in an energy interval $\Delta\subset
\Le(\tfrac{1}{2}B+\|V_\omega\|_\infty,\tfrac{3}{2}B-\|V_\omega\|_\infty\Ri)$
consists in the union of two sets $\Sigma_\ell$ and $\Sigma_r$,
which are small perturbations of the spectra
$\sigma(H_L+U_\ell+V_\omega^\ell)$ and
$\sigma(H_L+U_r+V_\omega^r)$, of the two single-boundary random Hamiltonians (see
Section \ref{sec2} for their precise definition). As in \cite{FM1}, the
eigenvalues in $\Sigma_{\ell}$ and $\Sigma_{r}$ are characterised by their
average velocity along the periodic direction $J_{E}=(\psi_E,v_y \psi_E)$: 
the eigenfunctions corresponding to the eigenvalues in
$\Sigma_\ell$ (resp. $\Sigma_r$) have a uniformly, negative (resp. positive) velocity,  with respect
to $L$. These are the so-called edge states and from the constructions in the proofs it is possible to see that the eigenvalues
in $\Sigma_{\ell}$ (resp. $\Sigma_{r}$) correspond to eigenfunctions localised
in the $x-$ direction near the left
(resp. right) boundary.  

Although our analysis is presented for a sample of size $L\times L$ the same results can be straightforwardly extended to all geometries
where the two boundaries are separated by any distance $D$ at least $O(\ln L)$ (assuming the length of the periodic direction is fixed to $L$). For 
distances $D=O(1)$ our analysis does not hold, a fact which is consistent with \cite{CHS}. In fact, we expect that by using
the results in the present paper one could prove that a wave packet localised on the left boundary and with appropriate energy,
will propagate along the left boundary up to a finite tunneling time and then, backscatter and propagate along the right boundary and so
forth. The tunneling time is set by $V_\omega$ and the distance $D$ between the two boundaries.
Thus if  $D=O(1)$ with respect to $L$, this tunneling time is also $O(1)$, and always remains much smaller than $O(L)$ which is the time needed for a ballistic flight around the whole periodic
direction $y$.

The paper is organised as follows. In section 2 we present the precise definition of the model and state the main Theorem.
 Section 3 is concerned with the main mathematical tools used in our analysis: a Wegner estimate and a decoupling scheme
 of the cylinder into two semi-infinite ones. 
 The proof of the main theorem is then completed in section 4. Some useful estimates and more technical material are collected in the appendices.

\section{The Model and Main Result}\label{sec2}
We study the spectral properties of the family of random
Hamiltonians
\begin{equation}\label{h}
H_\omega= H_L + U_\ell + U_r + V_\omega\, , \quad \omega \in
\Omega_\Lambda
\end{equation}
acting in the Hilbert space
$L^2(\R\times[-\tfrac{L}{2},\tfrac{L}{2}])$ with periodic boundary
conditions along $y$: $\psi(x,-\frac{L}{2}) =\psi(x,\frac{L}{2})$.
We choose the Landau gauge in which the kinetic part has the form
$H_L=\tfrac{1}{2}p_x^2 + \tfrac{1}{2}(p_y-Bx)^2$ with spectrum
given by the Landau levels: $\sigma(H_L)=\Le\{(n+\tfrac{1}{2})B;
n\in \N\Ri\}$. The potentials $U_\ell$ and $U_r$ representing the
confinement along the $x-$direction at $x=\pm \frac{L}{2}$ are
independent of $y$ and are supposed strictly monotonic, twice
differentiable and satisfy
\begin{eqnarray}
c_1|x+\tfrac{L}{2}|^{m_1}\leq U_\ell(x) \leq c_2|x+\tfrac{L}{2}|^{m_2} &\quad& \textrm{for  }x\leq
-\tfrac{L}{2} \label{U1}\\
c_1|x-\tfrac{L}{2}|^{m_1}\leq U_r(x) \leq c_2|x-\tfrac{L}{2}|^{m_2} &\quad&
\textrm{for  }x\geq
\tfrac{L}{2} \label{U2}\;
\end{eqnarray}
for some constants $0<c_1<c_2$, $2\leq m_1 < m_2<\infty$ and
$U_\ell(x)=0$ for $x\geq -\tfrac{L}{2}$, $U_r(x)=0$ for $x\leq
\tfrac{L}{2}$. \noindent The random potential $V_\omega$ is given
by the sum of local perturbations located at the sites of a finite
lattice $\Lambda=\Le\{(n,m)\in \Z^2 ; n\in
[-\frac{L}{2},\frac{L}{2}], m\in [-\frac{L}{2},\frac{L}{2}]
\Ri\}$. Let $V\geq 0$, with $V\in C^2$, $\|V\|_\infty \leq V_0$,
$\supp V\subset \mathbb{B}(\bs{0},\frac{1}{4})$ (the open ball
centered at $(0,0)$ of radius $\frac{1}{4}$) and $X_{n,m}(\omega)$ i.i.d. random variables with common bounded
density \mbox{$h \in C^2([-1,1])$} representing
the random strength of each local perturbation. Then $V_\omega$
has the form
\begin{equation}\label{randompot}
V_\omega(x,y)=\sum_{(n,m)\in \Lambda} X_{n,m}(\omega) V(x-n,y-m)
\end{equation}
We denote by
$\mathbb{P}_\Lambda$ the product measure defined on the set of all
possible realizations $\Omega_{\Lambda}=[-1,1]^\Lambda$. Clearly
for each realization $\omega\in \Omega_{\Lambda}$ we have
$\|V_\omega\|\leq V_0$ and we suppose
$V_0\ll B$. \\

For future use we collect some properties of three simpler random
Hamiltonians. \noindent Let us first consider the pure single-boundary Hamiltonians
\begin{equation}\label{3}
H_\alpha^0=H_L + U_\alpha \qquad\qquad \alpha=\ell, r \; .
\end{equation}
From translation invariance along $y$ we deduce that for
$L=+\infty$ the spectrum consists of analytic and monotone
decreasing (resp. increasing) branches
$\varepsilon_n^\ell({k})$ (resp. $\varepsilon_n^r({k})$)
where ${k}\in \R$ is the wave number associated to $p_y$.
One has $\lim_{{k}\to+\infty} \varepsilon_n^\ell({k})=
\lim_{{k}\to-\infty}
\varepsilon_n^r({k})=(n+\frac{1}{2})B$ and
$\lim_{k\to-\infty} \varepsilon_n^\ell({k})=
\lim_{{k}\to+\infty} \varepsilon_n^r({k})=+\infty$.
Because of periodic boundary conditions along $y$ the quantum
number ${k}$ takes discrete values $\frac{2\pi m}{L}$, $m\in
\Z$. For $L$ finite the spectrum consists of discrete eigenvalues
$E^\alpha_{n, m}=\varepsilon_n^\alpha(\frac{2\pi
m}{L})$ on the spectral branches. Moreover we have
\begin{equation}\label{dist}
\Le|E^\alpha_{0,m+1} - E^\alpha_{0,m}\Ri| \geq
\frac{C_0}{L} \qquad
\alpha=\ell, r
\end{equation}
for each $m$ such that $E_{0,m}^\alpha\in
\Delta_\varepsilon=\Le(\tfrac{1}{2}B+V_0+\varepsilon,\tfrac{3}{2}B-V_0-\varepsilon\Ri)$,
where $C_0>0$ is independent of $m$ and depends only on the spectral branch
$\varepsilon^\alpha_0$. We will suppose that the following
hypothesis is fulfilled
\begin{hyp}\label{hyp1}
There exists $L_0$ and $d_0>0$ such that for all $L>L_0$
\begin{equation}\label{hypot1}
\dist \Le(\sigma(H_\ell^0)\cap
\Delta_\varepsilon,\sigma(H_r^0)\cap \Delta_\varepsilon\Ri)\geq
\frac{d_0}{L} \; .
\end{equation}
\end{hyp}

In  order to fulfill this hypothesis one must take non-symmetric boundary potentials $U_\ell$ and $U_r$.
We expect that in fact our result still holds for $U_\ell(x)=U_r(-x)$ because physicaly the random potential $V_\omega$
removes with high probability any degeneracy, but in order to control this case one should improve the Wegner estimate
in Section 3. In Appendix C we give an example for a situation where this hypothesis is satisfied.

We will make use of the random single-boundary Hamiltonians
\begin{equation}\label{2}
H_\alpha=H_L+U_\alpha + V_\omega^\alpha
\end{equation}
where $V_\omega^\alpha=V_\omega|_{\Lambda_\alpha}$ with
$\Lambda_r=\Le\{(n,m)\in \Z^2 ; n\in
[\frac{L}{2}-\frac{3D}{4}-1,\frac{L}{2}], m\in
[-\frac{L}{2},\frac{L}{2}] \Ri\}$ and $\Lambda_\ell=\Le\{(n,m)\in
\Z^2 ; n\in [-\frac{L}{2},-\frac{L}{2}+\frac{3D}{4}+1], m\in
[-\frac{L}{2},\frac{L}{2}] \Ri\}$, where $D=\sqrt{L}$. \noindent
Since the perturbation has compact support and the essential
spectrum of $H_\alpha^0$ is given by the Landau levels, the
spectrum of $H_\alpha$ is discrete with the Landau levels as only
accumulation points. We denote it by
$\sigma(H_\alpha)=\Le\{E_\kappa^\alpha : \kappa \in \N\Ri\}$. One
can prove \cite{M} that, for each $\omega \in
\Omega_{\Lambda_\alpha}=[-1,1]^{\Lambda_\alpha}$ (the restriction
of the configurations $\omega$ to the sublattice $\Lambda_\alpha$)
and for each $\kappa$ such that $E^\alpha_{\kappa}\in
\Delta=(B-\delta,B+\delta)\subset \Delta_\varepsilon$ the distance
between two consecutive eigenvalues satisfies
\begin{equation}\label{distM}
\Le|E^\alpha_{\kappa+1} - E^\alpha_{\kappa}\Ri| \geq \frac{C}{L} \qquad\qquad
\alpha=\ell, r
\end{equation}
where $C>0$ is uniform in $\kappa$, $\omega$ and $L$. Moreover for each
$E^\ell_{\kappa}\in \Delta$ (resp. $E^r_{\kappa}\in \Delta$) the
average velocity associated to the corresponding
eigenfunctions is strictly negative (resp. positive) uniformly in
$L$ (see Appendix \ref{appB})
\begin{equation}\label{currentM}
\Le|J_{E^\alpha_\kappa}\Ri| \geq C'>0 \qquad\qquad
\alpha=\ell, r \; .
\end{equation}

Finally we remark that the Hamiltonian
$H_L+V_\omega|_{\tilde{\Lambda}}$ ($\tilde{\Lambda} \subset
\Lambda$) has a point spectrum contained in Landau bands
\begin{equation}
\sigma(H_L+V_\omega|_{\tilde{\Lambda}})\subset\bigcup_{n\geq 0}
\Le[(n+\tfrac{1}{2})B - V_0,(n+\tfrac{1}{2})B +V_0\Ri] \; .
\end{equation}
When $\tilde{\Lambda}$ is given by
$$\Lambda_b\equiv \tilde{\Lambda}=\Le\{(n,m)\in \Z^2 ; n\in
[-\tfrac{L}{2}+(\tfrac{D}{4}-1),\tfrac{L}{2}-(\tfrac{D}{4}-1)], m\in
[-\tfrac{L}{2},\tfrac{L}{2}] \Ri\}$$
we call the Hamiltonian $H_L+V_\omega|_{\Lambda_b}$ the bulk Hamiltonian and we denote it by $H_b$.
All the Hamiltonians considered so far are densely defined self-adjoint operators.\\

We now state the main result of this paper.

\begin{thm}\label{thm1}
Let $V_0$ small enough, fix $\varepsilon>0$ and let
$0<\delta<\frac{B}{2}-V_0-\varepsilon$. Suppose that $(H1)$ hold.
Then there exists $\mu>0$, $\bar{L}$ such that if $L>\bar{L}$ one
can find a set $\hat{\Omega}\subset \Omega_\Lambda$ of
realizations of the random potential $V_\omega$ with
$\mathbb{P}_\Lambda(\hat{\Omega})\geq 1-L^{-\nu}$ $(\nu\gg 1)$
such that for all $\omega\in \hat{\Omega}$ the spectrum of
$H_\omega$ in $\Delta=(B-\delta,B+\delta)$ is the union of two
sets $\Sigma_\ell$ and $\Sigma_r$ with the following properties:
\begin{itemize}
\item[a)] ${\cal E}_\kappa^\alpha \in \Sigma_\alpha$ $(\alpha=\ell,r)$ are a
small perturbation of $E_\kappa^\alpha\in \sigma(H_\alpha)\cap \Delta$ with
\begin{equation}
|{\cal E}_\kappa^\alpha-E_\kappa^\alpha| \leq e^{-\mu\sqrt{B}\sqrt{L}} \; .
\end{equation}
\item[b)] For ${\cal E}_\kappa^\alpha \in \Sigma_\alpha$ the average velocity $J_{{\cal
E}_\kappa^\alpha}$ of the associated eigenstate satisfies
\begin{equation}
|J_{{\cal E}_\kappa^\alpha}-J_{E_\kappa^\alpha}| \leq e^{-\mu\sqrt{B}\sqrt{L}}
\; .
\end{equation}
\end{itemize}
That is the eigenfunctions associated to the eigenvalues $($of
$H_\omega)$ in $\Delta$ have an ${\cal O}(1)$ velocity.
\end{thm}

The main tools for the proof of Theorem 1 are developed in section
\ref{sec3}. Basically they consist in a Wegner estimate for the
random Hamiltonians $H_\alpha$ ($\alpha=\ell,r$) and a decoupling
scheme that links the resolvent of the full Hamiltonian $H_\omega$
with those of $H_\ell$, $H_r$ and $H_b$. In section \ref{sec4} we
prove two propositions that lead to parts $a)$ and $b)$ of Theorem
1. Finally in appendix A we prove some technical results, in
appendix B we prove \eqref{currentM} and in appendix C we discuss
the Hypothesis 1.

Let $\bs{x}, \bs{x}'\in \R\times \Le[-\tfrac{L}{2},\tfrac{L}{2}\Ri]$, then one
can check that
\begin{eqnarray}
|\bs{x}-\bs{x}'|_\star \equiv \inf_{n\in \Z}\sqrt{(x-x')^2+(y-y'-nL)^2}
\end{eqnarray}
has the properties of a distance on $\R\times \mathbb{S}_L$ and that it is
related to the Euclidian distance $|\bs{x}-\bs{x}'| \equiv
\sqrt{(x-x')^2+(y-y')^2}$ by
\begin{equation}
|\bs{x}-\bs{x}'|_\star \leq |\bs{x}-\bs{x}'| \; .
\end{equation}
\noindent
The interest of $|\cdot|_\star$ is that, since we are working with a
cylindrical geometry all decay estimates are naturally expressed in terms of this
distance.

\section{Wegner Estimates and Decoupling Scheme}\label{sec3}

We first give a Wegner estimate for the Hamiltonians $H_\alpha$
($\alpha=\ell,r$). Denote by
$P_{0,m}^\alpha$ the projector of $H_\alpha^0$ onto the eigenvalue
$E_{0,m}^\alpha$ and by $P_\alpha(I)$ the projector of $H_\alpha$
on an interval $I$. Let $I_m=\Le(E_{0,
m-1}^\alpha+\delta_0,E_{0,
m}^\alpha-\delta_0\Ri)$ and $\Delta_\alpha=
\bigcup_{{m}_0\leq m \leq {m}_1} I_m$, for some
$-\infty \ll {m}_0<{m}_1 \ll \infty$ and
$\delta_0\ll\tfrac{C_0}{L}$. The local potentials $V(x-n,y-m)$
will also be denoted by $V_{\bs{i}}$, $\bs{i}=(m,n)\in \Lambda$.

\begin{prop}\label{prop1}
Let $V_0$ sufficiently small with respect to $B$, $E\in \Delta_\alpha\cap
\Delta_\varepsilon$ and  $I=[E-\bar{\delta},E+\bar{\delta}]\subset I_m$. Then
\begin{equation}
\mathbb{P}_{\Lambda_\alpha}\Le\{\dist(\sigma(H_\alpha),E)<\bar{\delta}\Ri\}
\leq \|h\|_{\infty}\bar{\delta} \dist(I,E_{0,\bar{m}}^\alpha)^{-2} V_0^2 L^4
\end{equation}
where $E_{0 \bar{m}}^\alpha$ is the closest eigenvalue of $\sigma(H_\alpha^0)$ to
the interval $I$.
\end{prop}

\begin{proof}
We first observe that $V_{\bs{i}}^{1/2}P_{0,m}^\alpha V_{\bs{j}}^{1/2}$ is trace class.
Indeed, using $\|AB\|_i\leq \|A\|\|B\|_i$ ($i=1,2$) and $\|AB\|_1\leq \|A\|_2\|B\|_2$
we get
$\|V_{\bs{i}}^{1/2}P_{0,m}^\alpha V_{\bs{j}}^{1/2}\|_1\leq
\|V_{\bs{i}}^{1/2}P_{0,m}^\alpha\|_2 \|P_{0,m}^\alpha V_{\bs{j}}^{1/2}\|_2 \leq
V_0\|P_{0,m}^\alpha\|_1^2 \leq V_0$.

We have $E\in \Delta_\alpha\cap \Delta_\varepsilon$, and $I=[E-\bar{\delta},E+\bar{\delta}]$ for
$\bar{\delta}$ small enough (we require that $I\subset \Delta_\alpha\cap
\Delta_\varepsilon$).
By the Chebyshev inequality we have
\begin{eqnarray}
\mathbb{P}_{\Lambda_\alpha}\Le\{\dist(\sigma(H_\alpha),E)<\bar{\delta}\Ri\} =
\mathbb{P}_{\Lambda_\alpha}\Le\{\Tr P_\alpha(I)\geq 1 \Ri\}\leq
\mathbb{E}_{\Lambda_\alpha}\{\Tr P_\alpha(I)\}
\end{eqnarray}
where $\mathbb{E}_{\Lambda_\alpha}$ is the expectation with respect to the random
variables in $\Lambda_\alpha$.

We first give an estimate on $\Tr P_\alpha(I)$. Let
$E^\alpha_{0,\bar{m}}$ the closest eigenvalue of
$\sigma(H_\alpha^0)$ to $I$ and $m_i$ ($i=0,1$) s.t.
$\dist(E^\alpha_{0,\bar{m}},E^\alpha_{0,m_i})={\cal O}(B)$. Let also
$P_>^\alpha=\sum_{m>m_1} P_{0,m}^\alpha$ and
$P_<^\alpha=\sum_{m<m_0}
P_{0,m}^\alpha$.\\
Using $P^\alpha_>(H^0_\alpha-E)P^\alpha_> \geq 0$ and
$P^\alpha_>R^0_\alpha(E)P^\alpha_>\leq
\dist(E^\alpha_{0,m_1+1},E)^{-1}P^\alpha_>$ we can
write
\begin{eqnarray}
P_\alpha(I) P^\alpha_> P_\alpha(I) &=& P_\alpha(I)
P^\alpha_>(H_\alpha^0-E)^{1/2}R^0_\alpha(E)(H_\alpha^0-E)^{1/2}P^\alpha_>
P_\alpha(I) \\
&\leq& \dist(E^\alpha_{0,m_1+1},E)^{-1}
\Le[P_\alpha(I) (H_\alpha-E)P^\alpha_>P_\alpha(I) - P_\alpha(I)
V_\omega^\alpha P^\alpha_> P_\alpha(I)\Ri] \nonumber
\end{eqnarray}
and thus
\begin{equation}\label{pqp12}
\|P_\alpha(I) P^\alpha_> P_\alpha(I)\| \leq
\dist(E^\alpha_{0,m_1+1},E)^{-1} \Le(\tfrac{|I|}{2}+
V_0\Ri) \leq \tfrac{1}{4}
\end{equation}
if, as we can suppose, $V_0$ is sufficiently small
($\dist(E^\alpha_{0,m_1+1},E)^{-1}V_0={\cal
O}\Le(\tfrac{V_0}{B}\Ri)$). \noindent In a similar way we get
\begin{equation}\label{pqp1}
\|P_\alpha(I) P^\alpha_< P_\alpha(I)\| \leq
\dist(E^\alpha_{0,m_0-1},E)^{-1} \Le(\tfrac{|I|}{2}+
V_0\Ri) \leq \tfrac{1}{4} \; .
\end{equation}
\noindent Now
\begin{eqnarray}
\Tr P_\alpha(I)P^\alpha_< = \Tr P_\alpha(I)P^\alpha_<P_\alpha(I)
\leq \|P_\alpha(I)P^\alpha_<P_\alpha(I)\| \Tr P_\alpha(I)
\end{eqnarray}
and similarly for $\Tr P_\alpha(I)P^\alpha_>$. Therefore, using
$1=P^\alpha_<+ P^\alpha_> + \sum_{m_0\leq m\leq m_1}
P_{0,m}^\alpha$, together with \eqref{pqp12} and \eqref{pqp1} we
obtain
\begin{equation}\label{cf}
\Tr P_\alpha(I)\leq 2\sum_{m_0\leq m\leq m_1}\Tr P_\alpha(I)
P_{0,m}^\alpha P_\alpha(I) \ \; .
\end{equation}
Since
\begin{equation}
\dist(I,E_{0,m}^\alpha)^2P_\alpha(I)^2 \leq
\Le(P_\alpha(I)(H_\alpha-E_{0,m}^\alpha)P_\alpha(I)\Ri)^2
\end{equation}
and $\dist(I,E_{0,m}^\alpha)^{-1}\leq
\dist(I,E_{0,\bar{m}}^\alpha)^{-1}$ for all $m_0\leq m\leq m_1$, it
follows that
\begin{eqnarray}\label{eqq2}
\Tr P_{0,m}^\alpha P_\alpha(I) P_{0,m}^\alpha &\leq&
\dist(I,E_{0,\bar{m}}^\alpha)^{-2} \times \nonumber \\
&\times&\Tr (P_{0k}^\alpha
P_\alpha(I)(H_\alpha-E_{0,m}^\alpha)P_\alpha(I)(H_\alpha-E_{0,m}^\alpha)P_\alpha(I)
P_{0,m}^\alpha)\nonumber \\
&=& \dist(I,E_{0,\bar{m}}^\alpha)^{-2}\Tr (P_{0,m}^\alpha
V_\omega^\alpha P_\alpha(I) V_\omega^\alpha P_{0,m}^\alpha) \; .
\end{eqnarray}
Thus, taking the expectation value in \eqref{cf} and using that
there are ${\cal O}(L)$ $m$'s between $m_0$ and $m_1$, we get
\begin{equation}\label{iin}
\mathbb{E}_{\Lambda_{\alpha}}\{\Tr P_\alpha(I)\} \leq 2\cdot {\cal O}(L) \cdot
\dist(I,E_{0,\bar{m}}^\alpha)^{-2}\sup_{m_0\leq m\leq m_1}\mathbb{E}_{\Lambda_{\alpha}}
\{\Tr (P_{0,m}^\alpha V_\omega^\alpha
P_\alpha(I)V_\omega^\alpha
P_{0,m}^\alpha)\} \; .
\end{equation}
It remains to estimate the expectation value in the right hand
side of \eqref{iin}. Here we follows a method of Combes and Hislop
\cite{CH}. Writing $V_{\omega}^\alpha=\sum_{\bs{i}\in
{\Lambda_\alpha}} X_{\bs{i}}(\omega)V_{\bs{i}}$
\begin{eqnarray}\label{33}
\Tr P_{0,m}^\alpha V_\omega^\alpha P_\alpha(I) V_\omega^\alpha P_{0,m}^\alpha &=&
\sum_{\bs{i},\bs{j} \in \Lambda_\alpha^2} X_{\bs{i}}(\omega)X_{\bs{j}}(\omega) \Tr P_{0,m}^\alpha
V_{\bs{i}}P_\alpha(I) V_{\bs{j}} P_{0,m}^\alpha \label{eq3} \\
&=&\sum_{\bs{i},\bs{j} \in \Lambda_\alpha^2} X_{\bs{i}}(\omega)X_{\bs{j}}(\omega)
\Tr V_{\bs{j}}^{1/2} P_{0,m}^\alpha
V_{\bs{i}}^{1/2} V_{\bs{i}}^{1/2} P_\alpha(I) V_{\bs{j}}^{1/2}\; .   \nonumber
\end{eqnarray}
Since $V_{\bs{j}}^{1/2}P_{0,m}^\alpha V_{\bs{i}}^{1/2}$ is trace
class we can introduce the singular value decomposition
\begin{equation}
V_{\bs{j}}^{1/2}P_{0,m}^\alpha V_{\bs{i}}^{1/2} =\sum_{n=0}^\infty\mu_n(u_n,.)v_n
\end{equation}
where $\sum_{n=0}^\infty\mu_n=\|V_{\bs{j}}^{1/2}P_{0,m}^\alpha V_{\bs{i}}^{1/2}\|_1$. Then
\begin{eqnarray}\label{44}
& &\Tr V_{\bs{j}}^{1/2}P_{0k}^\alpha
V_{\bs{i}}^{1/2} V_{\bs{i}}^{1/2}  P_\alpha(I) V_{\bs{j}}^{1/2} =\sum_{n=0}^\infty\mu_n
(u_n,V_{\bs{i}}^{1/2} P_\alpha(I) V_{\bs{j}}^{1/2}v_n)\nonumber
\\
&\leq& \sum_{n=0}^\infty\mu_n(v_n, V_{\bs{j}}^{1/2} P_\alpha(I) V_{\bs{j}}^{1/2}v_n)^{1/2}
(u_n, V_{\bs{i}}^{1/2} P_\alpha(I) V_{\bs{i}}^{1/2} u_n)^{1/2} \nonumber\\
&\leq&
\tfrac{1}{2}\sum_{n=0}^\infty \mu_n  \Le\{(v_n,V_{\bs{j}}^{1/2} P_\alpha(I)
V_{\bs{j}}^{1/2}v_n)+
(u_n,V_{\bs{i}}^{1/2} P_\alpha(I) V_{\bs{i}}^{1/2}u_n)\Ri\}  \; .
\end{eqnarray}
An application of the spectral averaging theorem (see \cite{CH}) shows that
\begin{equation}\label{spav}
\mathbb{E}_{\Lambda_\alpha}\{(v_n,V_{\bs{j}}^{1/2} P_\alpha(I) V_{\bs{j}}^{1/2}v_n)\}
\leq \|h\|_{\infty}2\bar{\delta}
\end{equation}
as well as for the term with $\bs{j}$ replacing $\bs{i}$ and $v_n$ replacing
$u_n$.
Combining \eqref{iin}, \eqref{44}, \eqref{spav} and \eqref{eq3} we get
\begin{eqnarray}
\mathbb{E}_{\Lambda_\alpha}\{\Tr P_\alpha(I)\}
&\leq& 4\cdot {\cal O}(L) \cdot \|h\|_\infty \bar{\delta} \dist(I,E_{0,\bar{m}}^\alpha)^{-2}
V_0^2 \sum_{\bs{i},\bs{j}\in \Lambda_\alpha^2} \|V_{\bs{j}}^{1/2}P_{0,m}^\alpha V_{\bs{i}}^{1/2}\|_1
\nonumber \\
&\leq& 4\cdot {\cal O}(L) \cdot \|h\|_\infty \bar{\delta} \dist(I,E_{0,\bar{m}}^\alpha)^{-2}
V_0^2|\Lambda_\alpha|^2 \; .
\end{eqnarray}
\end{proof}

We now turn to the decoupling scheme. By a decoupling formula \cite{BG},
\cite{BCD} the resolvent $R(z)=(z-H_\omega)^{-1}$ can be expressed, up to a
small term, as the sum of $R_\alpha(z)=(z-H_\alpha)^{-1}$ ($\alpha=\ell,r$) and
$R_b(z)=(z-H_b)^{-1}$. We set $D =\sqrt{L}$ and introduce the characteristic functions
\begin{eqnarray}
\tilde{J}_\ell(x)&=&\chi_{]-\infty,\mbox{\tiny$-\tfrac{L}{2}+\tfrac{D}{2}$}]}(x) \qquad
\tilde{J}_b(x)=\chi_{[\mbox{\tiny$-\tfrac{L}{2}+\tfrac{D}{2}$},
\mbox{\tiny$\tfrac{L}{2}-\tfrac{D}{2}$}]}(x)
\nonumber\\
\tilde{J}_r(x)&=&\chi_{[\mbox{\tiny$\tfrac{L}{2}-\tfrac{D}{2}$},+\infty[}(x) \;
.
\end{eqnarray}
We will also use three bounded $C^\infty(\R)$ functions $|J_i(x)|\leq 1$, $i\in {\cal
I}\equiv \{\ell,b,r\}$, with bounded first and second
derivatives $\sup_{x}|\partial_x^nJ_i(x)|\leq 2$, $n=1,2$,
and such that
\begin{eqnarray}
J_\ell(x)&=&
\begin{cases}
1 \quad \text{if } x\leq -\tfrac{L}{2}+\tfrac{3D}{4} \\
0 \quad \text{if } x\geq -\tfrac{L}{2}+\tfrac{3D}{4}+1
\end{cases} \qquad
J_b(x)=
\begin{cases}
1 \quad \text{if }  |x|\leq \tfrac{L}{2}-\tfrac{D}{4} \\
0 \quad \text{if }  |x|\geq \tfrac{L}{2}-\tfrac{D}{4}+1
\end{cases}
\nonumber\\
J_r(x)&=&
\begin{cases}
1 \quad \text{if } x\geq \tfrac{L}{2}-\tfrac{3D}{4} \\
0 \quad \text{if } x\leq \tfrac{L}{2}-\tfrac{3D}{4}-1
\end{cases} \; .
\end{eqnarray}

\begin{figure}[!h]\label{dec2wall}
\begin{center}
\input{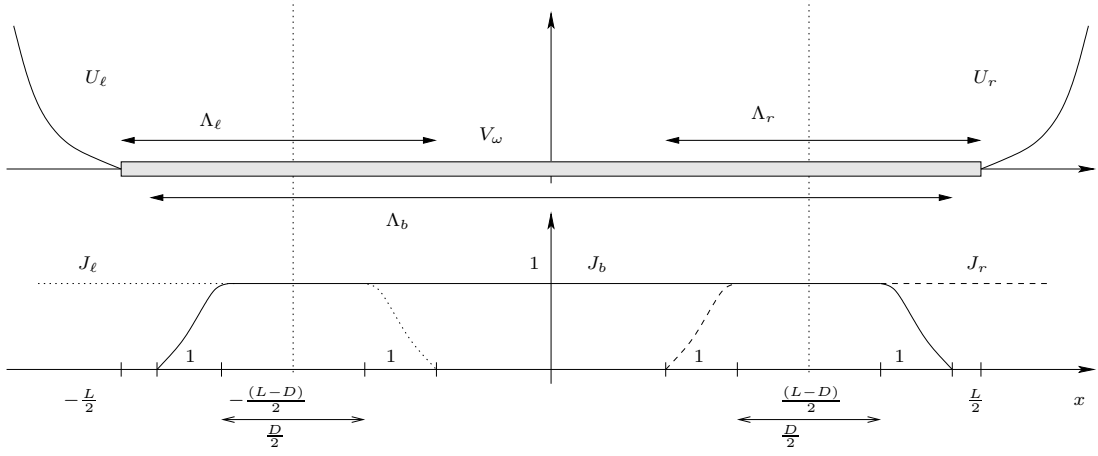}
\end{center}
\caption{\emph{The system of decoupling functions $J_i$ $(i\in {\cal I})$.}}
\end{figure}

For $i\in {\cal I}$ we have $H_\omega J_i=H_iJ_i$ and the
decoupling formula is \cite{BG}
\begin{equation}\label{resolvent}
R(z)=\Le(\sum_{i\in {\cal I}} J_i R_i(z) \tilde{J}_i\Ri)\Le( 1 -
{\cal K}(z)\Ri)^{-1}
\end{equation}
where
\begin{equation}
{\cal K}(z)=\sum_{i\in {\cal I}}K_i(z)=\sum_{i\in {\cal I}}
\tfrac{1}{2}[p_x^2,J_i]R_i(z)\tilde{J}_i \; .
\end{equation}
\noindent

The main result of this part is a lemma about $\|{\cal K}(z)\|$
for $z$ such that $\dist(z,\sigma(H_\alpha))\geq
e^{-\bar{\mu}\sqrt{B}\sqrt{L}}$, for a suitable $\bar{\mu}>0$ and
$\dist(z,\sigma(H_b))\geq \varepsilon$.

\begin{prop}\label{prop2}
Let $\varepsilon>0$, and $z\in \Delta_\varepsilon$ such that
$\dist(z,\sigma(H_\ell)\cup\sigma(H_r))\geq
e^{-\bar{\mu}\sqrt{B}\sqrt{L}}$ with $\bar{\mu}<\frac{1}{192}$.
Then for $L$ large enough there exists $C(B,V_0,\varepsilon)>0$
and $\tilde{\gamma}>0$ independent of $L$ such that
\begin{equation}\label{kappaz}
\|{\cal K}(z)\| \leq C(B,V_0,\varepsilon) e^{-\tilde{\gamma}\sqrt{B}\sqrt{L}} \; .
\end{equation}
\end{prop}

\begin{proof}
Computing the commutator in the definition of $K_i(z)$ we have
\begin{eqnarray}
K_i(z)&=&- \tfrac{1}{2}(\partial_x^2J_i)R_i(z)\tilde{J}_i -
(\partial_xJ_i)\partial_xR_i(z)\tilde{J}_i \; .
\end{eqnarray}
Then
\begin{eqnarray}\label{c}
\| K_b(z)\| &\leq&  \tfrac{1}{2}\|(\partial^2_xJ_b)R_b(z)\tilde{J}_b\| +
\|(\partial_xJ_b)\partial_xR_b(z)\tilde{J}_b\|  \\
\| K_\alpha(z)\| &\leq&
\tfrac{1}{2}\|(\partial^2_xJ_\alpha)R_\alpha^b(z)\tilde{J}_\alpha\| +
\tfrac{1}{2}\|(\partial^2_xJ_\alpha)R_\alpha^b(z)U_\alpha\|\,\dist(z,\sigma(H_\alpha))^{-1} \\
&+& \|(\partial_xJ_\alpha)\partial_xR_\alpha^b(z)\tilde{J}_\alpha\|
+
\|(\partial_xJ_\alpha)\partial_xR_\alpha^b(z)U_\alpha\|\,\dist(z,\sigma(H_\alpha))^{-1}
\nonumber
\end{eqnarray}
where for the the second term we used the second resolvent identity and where
$R_\alpha^b(z)=(z-[H_L+V_\omega^\alpha])^{-1}$. \\
\noindent We have to estimate norms of the form
$\|f\partial_x^\alpha \tilde{R}(z)g\|$ ($\alpha=0,1$) where here
$\tilde{R}(z)$ is $R_b(z)$ or $R^b_\alpha(z)$, $f=\partial^m_xJ_i$
and $g=\tilde{J}_i$ or
$g=U_\alpha$.\\
Using the second resolvent formula we develop $\tilde{R}(z)$ in
its Neumann series, denote $V_\omega|_{\tilde{\Lambda}}\equiv W$
($\tilde{\Lambda}=\Lambda_b$ or $\Lambda_\alpha$)
\begin{equation}
\tilde{R}(z) = \sum_{n=0}^\infty R_0(z)[WR_0(z)]^n
\end{equation}
where $R_0(z)=(z-H_L)^{-1}$. The norm convergence is ensured since
we are in a spectral gap, indeed
\begin{equation}
\|WR_0(z)\| \leq V_0 \dist(z,\sigma(H_L))^{-1} \leq
\frac{V_0}{V_0+\varepsilon} < 1 \; .
\end{equation}
Therefore
\begin{equation}\label{dyson}
\|f\partial^\alpha_x\tilde{R}(z)g\| \leq \sum_{n=1}^\infty
\|f\partial^\alpha_x R_0(z)\Le[WR_0(z)\Ri]^n g\|
\end{equation}
and we have to control the operator norms $\|f\partial^\alpha_xR_0(z)\Le[WR_0(z)\Ri]^n
g\|$.\\
For any vector $\varphi \in
L^2(\R\times[-\frac{L}{2},\frac{L}{2}])$ with $\|\varphi\|=1$
\begin{equation}\label{1}
\|f\partial^\alpha_xR_0(z)\Le[WR_0(z)\Ri]^ng\varphi\|^2 =
\int_{\supp f} |f(\bs{x})|^2 \Le|(\partial_x^\alpha
R_0(z)\Le[WR_0(z)\Ri]^ng\varphi)(\bs{x})\Ri|^2 \D \bs{x}
\end{equation}
For the integrand in \eqref{1} we have
\begin{eqnarray}
{\cal J}&\equiv& \Le|(\partial_x^\alpha R_0(z)\Le[WR_0(z)\Ri]^ng\varphi)(\bs{x})\Ri| \leq
\int_{\supp g} \D \bs{x}' \int \D \bs{x}_1 \ldots \D \bs{x}_{n}
 \times \\
&\times&|\partial^\alpha_x R_0(\bs{x},\bs{x}_1;z)||W(\bs{x_1})||R_0(\bs{x}_1,\bs{x}_2;z)| \ldots
|W(\bs{x_n})||R_0(\bs{x}_{n},\bs{x}';z)||g(\bs{x}')||\varphi(\bs{x}')| \nonumber
\; .
\end{eqnarray}
Now, taking out $\|W\|_\infty$ and using Lemma \ref{lem9},
Appendix \ref{appA} we get
\begin{eqnarray}
{\cal J} &\leq& \Le(cB^2 \tfrac{V_0}{V_0+\varepsilon}\Ri)^n
\int_{\supp g} \D \bs{x}' \int \D \bs{x}_1 \ldots \D \bs{x}_{n}
e^{-\bar{\gamma}\sqrt{B} \sum_{i=0}^{n}
|\bs{x}_i-\bs{x}_{i+1}|_\star}\times \nonumber \\
&\times&|\Phi^1(|\bs{x}-\bs{x}_1|_\star)|
 \ldots
|\Phi^0(|\bs{x}_{n}-\bs{x}'|_\star)||g(\bs{x}')||\varphi(\bs{x}')|
\end{eqnarray}
where $\bs{x}_{0}=\bs{x}$ and $\bs{x}_{n+1}=\bs{x}'$. Splitting
the exponential and making the change of variables
$\bs{x}-\bs{x}_1=-\bs{z}_1$, \ldots,
$\bs{x}_{n-1}-\bs{x}_{n}=-\bs{z}_{n}$ we get (with
$\bs{x}_{n}=\bs{x}_{n}(\{\bs{z}_i\},\bs{x})$ and
$A=cB^2\frac{V_0}{V_0+\varepsilon}$)
\begin{eqnarray}
{\cal J} &\leq& A^n \sup_{\bs{z}_1 \ldots \bs{z}_{n}} \Le\{ \int_{\supp g}
e^{-\frac{2}{3}\bar{\gamma} \sqrt{B}|\bs{x}-\bs{x}'|_\star}
|g(\bs{x}')||\varphi(\bs{x}')||\Phi^0(|\bs{x}_{n}-\bs{x}'|_\star)|
e^{-\frac{1}{3}\bar{\gamma}\sqrt{B}|\bs{x}_{n}-\bs{x}'|_\star}\D \bs{x}'\Ri\} \times\nonumber \\
&\times& \Le[\int_{\R^2}
|\Phi^1(|\bs{z}|)|e^{-\frac{1}{3}\bar{\gamma}\sqrt{B}|\bs{z}|} \D
\bs{z}\Ri] \Le[\int_{\R^2}
|\Phi^0(|\bs{z}|)|e^{-\frac{1}{3}\bar{\gamma}\sqrt{B}|\bs{z}|} \D
\bs{z}\Ri]^{n-1} \noindent \\ &\equiv& A^n \sup_{\bs{z}_1 \ldots
\bs{z}_{n}} \{{\cal X}\}\, [{\cal Y}] \,[{\cal Z}]^{n-1} \; .
\end{eqnarray}
Splitting the exponential and using the Schwartz inequality  we have
the estimate
\begin{eqnarray}
\sup_{\bs{z}_1 \ldots \bs{z}_{n}} {\cal X} &\leq& \sup_{\bs{x}'\in
\supp g}
e^{-\frac{1}{3}\bar{\gamma}\sqrt{B}|\bs{x}-\bs{x}'|_\star}
\Le\{\int_{\R^2}|\Phi^0(|\bs{w}|)|^2e^{-\frac{2}{3}\bar{\gamma}\sqrt{B}|\bs{w}|}
\D
\bs{w}\Ri\}^{1/2}\times \nonumber \\
&\times& \Le(\sup_{\bs{x}' \in \supp g}
e^{-\frac{2}{3}\bar{\gamma}\sqrt{B}|x-x'|} |g(x')|^2
\Ri)^{1/2}\|\varphi\| \; .
\end{eqnarray}
Now, since $U_\alpha$ do not grow to fast (see \eqref{U1},
\eqref{U2}) $(\sup_{\bs{x}' \in \supp g}
e^{-\frac{2}{3}\bar{\gamma}\sqrt{B}|x-x'|} |g(x')|^2)^{1/2}$ is
bounded by a numerical constant. On the other and the term
$\int_{\R^2}|\Phi^0(|\bs{w}|)|^2e^{-\frac{2}{3}\bar{\gamma}\sqrt{B}|\bs{w}|}
\D \bs{w}$ is bounded by a constant depending only on $B$.

Moreover the terms ${\cal Y}$ and ${\cal Z}$ are also bounded by a
constant depending only on $B$ and not on $L$. This leads to
\begin{equation}
\|f\partial^\alpha_x\Le[R_0(z)\Ri]^ng\varphi\| \leq \|f\|_\infty
\hat{C}(B) (\tilde{C}(B)A)^n e^{-\frac{1}{12}\bar{\gamma}\sqrt{B}
D} \|\varphi\| \; .
\end{equation}
Therefore, if $V_0$ is small enough the series \eqref{dyson}
converges and
\begin{equation}\label{po}
\|f\partial^\alpha_x\tilde{R}(z)g\| \leq
\tilde{C}(B,V_0)\sqrt{L}e^{-\frac{1}{12} \bar{\gamma}\sqrt{B} D}
\; .
\end{equation}
This implies
\begin{eqnarray}
\|K_b(z)\| &\leq& \varepsilon^{-1}\sqrt{L} C(B,V_0)e^{-\frac{1}{12}
\bar{\gamma}\sqrt{B} \sqrt{L}} \\
\|K_\alpha(z)\| &\leq& \sqrt{L} e^{\bar{\mu}\sqrt{B}\sqrt{L}}C(B,V_0)e^{-\frac{1}{12}
\bar{\gamma}\sqrt{B} \sqrt{L}} \qquad \alpha=\ell, r
\end{eqnarray}
thus $\|{\cal K}(z)\|\leq
C(B,V_0,\varepsilon)e^{-\tilde{\gamma}\sqrt{B} \sqrt{L}}$ where
$2\tilde{\gamma}=\frac{\bar{\gamma}}{12}-\bar{\mu}$. Since
$\bar{\gamma}=\frac{1}{16}$ in Lemma \ref{lem9}, Appendix A we
must take $\bar{\mu}<\frac{1}{192}$.
\end{proof}

We remark that in the proof above we have proved the
following statement (see \eqref{po}) that will be useful in the
next section
\begin{equation}\label{rem}
\|(1-\tilde{J}_\alpha)\tilde{R}_b(z)g\| \leq
\bar{C}(B,V_0,\varepsilon)e^{- \tilde{\gamma}\sqrt{B} \sqrt{L}} \;
.
\end{equation}
where $g=U_\alpha$ or $g=\chi_B$
($B\subset\R\times[-\tfrac{L}{2},\tfrac{L}{2}]$) with $\dist(\supp
g,\supp(1-\tilde{J}_\alpha)) = {\cal O}(D)$ and $\tilde{R}_b(z)$ a
resolvent associated to a generic bulk Hamiltonian
($H_L+V_\omega|_{\tilde{\Lambda}})$.

\section{Projector estimates and the proof of Theorem 1}\label{sec4}

In this section we prove two propositions that lead to Theorem 1.
Let ${\cal D}'=\{\kappa: E_\kappa^\alpha \in \Delta, \alpha=\ell,
r\}$, $\text{card}({\cal D}')={\cal O}(L)$, where $\Delta\subset
\Delta_\varepsilon$ is given in section 2.

\begin{prop}\label{prop3}
For $L$ large enough, with probability greater then $1-L^{-\nu}$ $(\nu \gg 1)$, we have for all
$\kappa \in {\cal D}'$
\begin{equation}
\|P-P_\alpha(E_\kappa^\alpha)\| \leq e^{-\gamma\sqrt{B}\sqrt{L}}
\end{equation}
where $P_\alpha(E_\kappa^\alpha)$ is the projector associated to $H_\alpha$ onto
$E_\kappa^\alpha$ and $P$ is the projector associated to $H_\omega$ onto $\{z\in
\C: |z-E_\kappa^\alpha|\leq e^{-\bar{\mu}\sqrt{B}\sqrt{L}}\}$.
\end{prop}

\begin{proof}
(1): Let ${\cal E}=\{m: E_{0,m}^\alpha \in \Delta, \alpha=\ell, r\}$,
$\text{card}({\cal E})={\cal O}(L)$, and let
\begin{equation}
\hat{\Omega}_\ell=\{\omega \in \Omega_{\Lambda_\ell} :
\dist(E_{0,m}^r,\sigma(H_\ell))\geq L^{-\sigma}, \forall m\in {\cal E}\} \; ,
\end{equation}
with $\sigma>11$, this set has probability
\begin{equation}
\mathbb{P}_{\Lambda_\ell}(\hat{\Omega}_\ell) \geq
1-L^{-(\sigma-8)} \; .
\end{equation}
Indeed for a fixed $m\in {\cal E}$, using Proposition
\ref{prop1} and $(H1)$ one gets
\begin{eqnarray}
& &\mathbb{P}_{\Lambda_\ell} \Le\{
\omega \in \Omega_{\Lambda_\ell} :
\dist(E_{0,m}^r,\sigma(H_\ell))\geq L^{-\sigma}, \text{ for one } m\in {\cal E}
\Ri\} \nonumber \\
&\geq& 1-C'(h,V_0)L^{-\sigma}L^4\Le(\tfrac{d_0}{L}-L^{-\sigma}
\Ri)^{-2} \geq 1-C(h,V_0)L^{6-\sigma} \; .
\end{eqnarray}
For a given realisation $\omega_\ell \in \hat{\Omega}_\ell$ let 
\begin{equation}
\hat{\Omega}_r(\omega_\ell)=\{\omega \in \Omega_{\Lambda_r} :
\dist(E_{\kappa}^\ell,\sigma(H_r))\geq L^{-3\sigma}, \forall \kappa\in {\cal D}'\}
\; ,
\end{equation}
this set has probability
\begin{equation}
\mathbb{P}_{\Lambda_r}(\hat{\Omega}_r(\omega_\ell)|\omega_\ell) \geq 1-L^{-(\sigma-6)} \; .
\end{equation}
uniformly with respect to the realisations of $\hat{\Omega}_\ell$.
Indeed
\begin{eqnarray}
& &\mathbb{P}_{\Lambda_r} \Le\{
\omega \in \Omega_{\Lambda_r} :
\dist(E_\kappa^\ell,\sigma(H_r))\geq L^{-3\sigma}, \text{ for one } \kappa\in
{\cal D}' \Ri\} \nonumber \\
&\geq& 1-C'(h,V_0)L^{-3\sigma}L^4\Le(L^{-\sigma}-L^{-3\sigma}\Ri)^{-2}
\geq 1-C(h,V_0)L^{4-\sigma} \; .
\end{eqnarray}
It follows that the set
\begin{equation}
\hat{\Omega}^{(\ell)} = \Le\{ \omega=(\omega_\ell,\omega_b,\omega_r) \in \Omega: \omega_\ell \in \hat{\Omega}_\ell, \omega_b
\in \Omega_b, \omega_r\in \hat{\Omega}_r(\omega_\ell) \Ri\}
\end{equation}
$\Omega_b=\Omega|_{\Lambda_b \backslash (\Lambda_\ell \cup
\Lambda_r)}$ has probability
\begin{eqnarray}
\mathbb{P}_\Lambda(\hat{\Omega}^{(\ell)}) &=& \mathbb{P}_{\Lambda_b}(\hat{\Omega}_b) \mathbb{E}_{\Lambda_\ell} \Le\{ \mathbb{P}_{\Lambda_r}
(\hat{\Omega}_r|\omega_\ell) \big|\omega_\ell \in \hat{\Omega}_\ell \Ri\} \nonumber \\
&\geq& (1-L^{-(\sigma-6)})\mathbb{P}_{\Lambda_\ell}(\hat{\Omega}_\ell) \geq
1-L^{-(\sigma-9)}
\end{eqnarray}

(2): We now work with a given $\omega \in \hat{\Omega}^{(\ell)}$. Take $\bar{\mu}>0$ as in
Proposition \ref{prop2} and $L$ large enough such that
for all $\kappa \in {\cal D}'$ $\Gamma_\kappa=\{z\in \C: |z-E_\kappa^\ell|\leq
e^{-\bar{\mu}\sqrt{B}\sqrt{L}}\}\cap \sigma(H_r)=\emptyset$, and remark that $\Tr
P_b(\Delta)=0$ ($P_b$ the projector associated to $H_b$). \\
We need to introduce two auxiliary Hamiltonians $H_1$ and $H_2$ defined as follows:
\begin{eqnarray}
H_1&=&H_L + V_\omega^\ell|_{\Lambda_1} \\
H_2&=&H_L + V_\omega^\ell|_{\Lambda_2} +U_\ell
\end{eqnarray}
where $\Lambda_2=
\Le\{(n,m)\in \Z^2 ; n\in
[-\frac{L}{2},-\frac{L}{2}+(\frac{D}{4}-1)], m\in [-\frac{L}{2},\frac{L}{2}] \Ri\}$,
and $\Lambda_1=\Lambda_\ell \backslash \Lambda_2$, of course
$H_\ell=H_2+V_\omega^\ell|_{\Lambda_1}$.

From the decoupling formula \eqref{resolvent} we have
\begin{eqnarray}
R(z)-R_\ell(z) &=&\Le(\sum_{i\in {\cal I}} J_iR_i(z)\tilde{J}_i\Ri)
\Le(\sum_{n=1}^\infty {\cal K}(z)^n\Ri)
-(1-J_\ell)R_\ell(z)\nonumber\\ &-& J_\ell R_\ell(z)(1-\tilde{J}_\ell)
+ J_b R_b(z)\tilde{J}_b + J_rR_r(z)\tilde{J}_r\; .
\end{eqnarray}
integrating over $\partial\Gamma_\kappa$ and taking the operator norm we get
\begin{eqnarray}\label{pdb}
\|P-P_\ell(E_\kappa^\ell)\| &\leq& e^{-\bar{\mu}\sqrt{B}\sqrt{L}} \Le(\sum_{i\in {\cal I}}\sup_{z\in
\partial\Gamma_\kappa}\|R_i(z)\|\Ri)
\frac{\sup_{z\in \partial\Gamma_\kappa} \|{\cal K}(z)\|}{1- \sup_{z\in \partial\Gamma_\kappa} \|{\cal K}(z)\|} \nonumber\\
&+& \|(1-J_\ell)P_\ell(E_\kappa^\ell)\|+\|J_\ell P_\ell(E_\kappa^\ell)(1-\tilde{J}_\ell)\|
\nonumber \\
&=& a + b + c\; .
\end{eqnarray}
For the first term we note that for $L$ large enough
$e^{-\bar{\mu}\sqrt{B}\sqrt{L}}\sup_{z\in \partial\Gamma_\kappa}\|R_i(z)\|\leq
1$ ($i\in {\cal I}$). Indeed, for $i=\ell$ we have $\sup_{z\in \partial\Gamma_\kappa}\|R_\ell(z)\|=
e^{\bar{\mu}\sqrt{B}\sqrt{L}}$ by construction, for $i=b$ we have
$\sup_{z\in \partial\Gamma_\kappa}\|R_b(z)\|=\varepsilon^{-1}$ and for $i=r$
$\sup_{z\in \partial\Gamma_\kappa}\|R_r(z)\|=\Le(L^{-3\sigma}-e^{-\bar{\mu}\sqrt{B}\sqrt{L}}\Ri)^{-1}$.
Then, applying Proposition \ref{prop2} we get
\begin{equation}
a \leq 2C(B,V_0,\varepsilon) e^{-\tilde{\gamma} \sqrt{B}
\sqrt{L}}\; .
\end{equation}
For the second and third term we first observe that by the second resolvent formula
\begin{equation}\label{forqw}
\frac{P_\ell(E^\ell_\kappa)}{(z-E^\ell_\kappa)}=(z-H_1)^{-1}P_\ell(E^\ell_\kappa) +
(z-H_1)^{-1}[V_\omega^\ell|_{\Lambda_2}+U_\ell]\frac{P_\ell(E^\ell_\kappa)}{(z-E^\ell_\kappa)} \; .
\end{equation}
and integrating \eqref{forqw} along $\partial\Gamma_\kappa$ we obtain
(using $\sigma(H_1)\cap \Delta_\varepsilon= \emptyset$)
\begin{eqnarray}
P_\ell(E_\kappa^\ell)&=&R_1(E_\kappa^\ell)[V_\omega^\ell|_{\Lambda_2}+U_\ell]P_\ell(E_\kappa^\ell)\label{forb}\\
&=&P_\ell(E_\kappa^\ell) [V_\omega^\ell|_{\Lambda_2}+U_\ell]
R_1(E_\kappa^\ell) \label{forc} \; .
\end{eqnarray}
Therefore, using \eqref{forb} for $b$ and \eqref{forc} for $c$ we get
\begin{eqnarray}
b &\leq& \|(1-J_\ell)R_1(E_\kappa^\ell)[V_\omega^\ell|_{\Lambda_2}+U_\ell]\|
\leq \|(1-\tilde{J}_\ell)R_1(E_\kappa^\ell)[V_\omega^\ell|_{\Lambda_2}+U_\ell]\|\\
c &\leq&
\|(1-\tilde{J}_\ell)R_1(E_\kappa^\ell)[V_\omega^\ell|_{\Lambda_2}+U_\ell]\|  \; .
\end{eqnarray}
Using \eqref{rem} we get
\begin{eqnarray}
b+c &\leq& 2\Le( V_0L^2
\|(1-\tilde{J}_\ell)R_1(E_\kappa^\ell)
\chi_{\Lambda_2}\| +\|(1-\tilde{J}_\ell)R_1(E_\kappa^\ell)U_\ell\| \Ri)\nonumber\\
&\leq& 2\bar{C}(B,V_0,\varepsilon) L^2e^{-\tilde{\gamma}\sqrt{B}
\sqrt{L}}\; .
\end{eqnarray}
Thus
\begin{equation}\label{dll}
\|P-P_\ell(E_\kappa^\ell)\| \leq e^{-\gamma\sqrt{B}\sqrt{L}} \; .
\end{equation}
By repeating the above proof in a symmetrical way we get for $\omega$ in a set $\hat{\Omega}^{(r)}$ similar 
to $\hat{\Omega}^{(\ell)}$
\begin{equation}\label{drr}
\|P-P_r(E_\kappa^r)\| \leq e^{-\gamma\sqrt{B}\sqrt{L}} \; .
\end{equation}
Finally we have both \eqref{dll} and \eqref{drr} for $\omega\in
\hat{\Omega}=\hat{\Omega}^{(\ell)} \cap \hat{\Omega}^{(r)}$ with
$\mathbb{P}_\Lambda\geq 1-L^{-\nu}$, $\nu=\sigma-10$.
Note that we can take $\nu'\gg 1$ by taking $\sigma \gg 11$.

\end{proof}
The estimate on the norm difference of the projectors implies that their dimensions are the same and that ${\cal
E}_\kappa^\alpha\in \sigma(H_\omega)$ is a small perturbation of
$E_\kappa^\alpha$: this gives part $a)$ of Theorem 1.

\begin{prop}\label{prop4}
Let $\omega \in \hat{\Omega}$. Then there exists $\hat{\mu}>0$
such that the velocity associated to each eigenvalue ${\cal
E}_\kappa^\alpha$ of $H_\omega$ in $\Delta$ satisfies
\begin{equation}
\Le|J_{{\cal E}_\kappa^\alpha}-J_{E_\kappa^\alpha}\Ri| \leq e^{-\hat{\mu} \sqrt{B} \sqrt{L}}
\; .
\end{equation}
\end{prop}

\begin{proof}
Let $J_{{\cal E}_\kappa^\alpha}=\Tr v_yP({\cal E}_\kappa^\alpha)$ the average velocity associated to the
eigenvalue ${\cal E}_\kappa^\alpha\in \sigma(H_\omega)$ and $J_{E_\kappa^\alpha}=\Tr
v_yP_\alpha(E_\kappa^\alpha)$
that associated to the eigenvalue $E_\kappa^\alpha$ of $H_\alpha$.
First we observe that
$v_yP({\cal E}_\kappa^\alpha)$ is trace class. Indeed,
$v_yP({\cal E}_\kappa^\alpha)=v_yP({\cal E}_\kappa^\alpha)P({\cal E}_\kappa^\alpha)$ with $v_yP({\cal E}_\kappa^\alpha)$ bounded and
$\|P({\cal E}_\kappa^\alpha)\|_1=\Tr P({\cal E}_\kappa^\alpha)= \Tr P_\alpha(E_\kappa^\alpha)= 1$.
\begin{eqnarray}\label{tn1}
\|v_yP({\cal E}_\kappa^\alpha)\|_1^2&\leq&\|v_yP({\cal E}_\kappa^\alpha)\|^2
\leq \|P({\cal E}_\kappa^\alpha) v_y^2 P({\cal E}_\kappa^\alpha)\|\\
&\leq&
2\|P({\cal E}_\kappa^\alpha) (H_\omega-V_\omega) P({\cal E}_\kappa^\alpha)\| \leq \Le(3B+2V_0\Ri)\nonumber
\end{eqnarray}
To get the second inequality one has simply added positive terms to
$v_y^2$.
Similarly
\begin{equation}\label{tn2}
\|v_y P_\alpha(E_\kappa^\alpha)\|_1^2\leq (3B+2V_0)\; .
\end{equation}
With the help of the identity
\begin{eqnarray}\label{iden}
P({\cal E}_\kappa^\alpha)-P_\alpha(E_\kappa^\alpha) &=&
[P({\cal E}_\kappa^\alpha)-P_\alpha(E_\kappa^\alpha)]^2 +
[P({\cal E}_\kappa^\alpha)-P_\alpha(E_\kappa^\alpha)]P_\alpha(E_\kappa^\alpha) \nonumber \\
&+& P_\alpha(E_\kappa^\alpha)[P({\cal E}_\kappa^\alpha)-P_\alpha(E_\kappa^\alpha)]
\end{eqnarray}
we get
\begin{eqnarray}\label{equal}
|J_{{\cal E}_\kappa^\alpha}-J_{E_\kappa^\alpha}|&=&
\Le|\Tr v_y[P({\cal E}_\kappa^\alpha)-P_\alpha(E_\kappa^\alpha)]\Ri|
\leq \Le|\Tr v_y[P({\cal E}_\kappa^\alpha)-P_\alpha(E_\kappa^\alpha)]^2\Ri| \nonumber\\
&+&\Le|\Tr v_y[P({\cal E}_\kappa^\alpha)-P_\alpha(E_\kappa^\alpha)]P_\alpha(E_\kappa^\alpha)\Ri|\nonumber \\
&+&\Le|\Tr v_yP_\alpha(E_\kappa^\alpha)[P({\cal E}_\kappa^\alpha)-P_\alpha(E_\kappa^\alpha)]\Ri| \; .
\end{eqnarray}
and then, from \eqref{tn1} and \eqref{tn2}, we get
\begin{eqnarray}
|J_{{\cal E}_\kappa^\alpha}-J_{E_\kappa^\alpha}| &\leq& 2 \Le(
\|v_yP({\cal E}_\kappa^\alpha)\|_1 + \|v_yP_\alpha(E_\kappa^\alpha)\|_1 \Ri)
\|P({\cal E}_\kappa^\alpha)-P_\alpha(E_\kappa^\alpha)\| \\
&\leq&
4(3B+2V_0)^{1/2} \|P({\cal E}_\kappa^\alpha)-P_\alpha(E_\kappa^\alpha)\|
\nonumber\; .
\end{eqnarray}
Combining this last inequality with Proposition \ref{prop3} we get the result.
\end{proof}

From Proposition \ref{prop4} and the result of Appendix \ref{appB} given in
\eqref{currentM} we obtain part $b)$ of Theorem 1.

\appendix

\section{Estimate of the Green function $R_0(\bs{x},\bs{x}';z)$}\label{appA}

In this appendix we give the necessary decay property of the kernel $R_0(\bs{x},\bs{x}';z)$
with periodic boundary conditions along $y$.
The exact formula for $R_0(\bs{x},\bs{x}';z)$ can be found in \cite{FM1}. We introduce the
following notation
\begin{eqnarray}
& &\Phi^\alpha(|\bs{x}-\bs{x}'|_\star) \nonumber\\
&=&
\begin{cases}
1+\Le|\ln
\Le(\tfrac{B}{2}|\bs{x}-\bs{x}'|_\star^2\Ri)\Ri| \quad , \; \alpha=0 \\
1+\Big[ \Le|\ln
\Le(\tfrac{B}{2}|\bs{x}-\bs{x}'|_\star^2\Ri)\Ri|
+\Le(1+\Le|\ln
\Le(\tfrac{B}{2}|\bs{x}-\bs{x}'|_\star^2\Ri)\Ri|\Ri)|\bs{x}-\bs{x}'|_\star^{-1}\Big]
\quad , \; \alpha=1 \; . \label{PHYa}
\end{cases}
\end{eqnarray}

\begin{lem}\label{lem9}
If $|{\cal I}m \,z|\leq 1$, ${\cal R}e \,z \in
\,\Le]\tfrac{1}{2}B,\tfrac{3}{2}B\Ri[$ then, for $L$ large enough, there exists
 $C(z,B)$ positive constant independent of $L$ such that
 $(\alpha=0,1)$
\begin{eqnarray}
|\partial_x^\alpha R_0(\bs{x},\bs{x}';z)|
&\leq& C'(z,B) e^{-\frac{B}{8}|\bs{x}-\bs{x}'|^2_\star}
\Phi^\alpha(|\bs{x}-\bs{x}'|_\star) \nonumber\\
&\leq&C(z,B) e^{-\bar{\gamma}\sqrt{B}|\bs{x}-\bs{x}'|_\star}\Phi^\alpha(|\bs{x}-\bs{x}'|_\star)
\end{eqnarray}
where $C(z,B)= cB^2\dist(z,\sigma(H_L))^{-1}$ with $c$ a numerical
positive constant and $\bar{\gamma}=\frac{1}{16}$.
\end{lem}

\begin{proof}
As in \cite{FM1} we can prove that (for $L$ large enough the
logarithmic divergences appear only for $|m|\leq 1$ and the sum
over $|m|>1$ converge)
\begin{equation} \label{ght}
|\partial_x^\alpha R_0(\bs{x},\bs{x}';z)| \leq
\tfrac{C'(z,B)}{3} e^{-\frac{B}{8}|\bs{x}-\bs{x}'|^2} + \sum_{|m|\leq 1}
|\partial_x^\alpha R_0^\infty(x\, y-mL,\bs{x}';z)|
\end{equation}
with
\begin{eqnarray}\label{qqq}
& &|\partial_x^\alpha R_0^\infty(\bs{x},\bs{x}';z)| \label{bu} \\
&\leq&
\begin{cases}
\tfrac{C'(z,B)}{3} e^{-\frac{B}{8}|\bs{x}-\bs{x}'|^2}
\Le\{1+\bs{1}_{\mathbb{B}(\bs{0},\sqrt{2B^{-1}})}(|\bs{x}-\bs{x}'|) \Le|\ln
\Le(\tfrac{B}{2}|\bs{x}-\bs{x}'|^2\Ri)\Ri|\Ri\}, \quad \alpha=0 \\
\tfrac{C'(z,B)}{3}e^{-\frac{B}{8}|\bs{x}-\bs{x}'|^2}
\Big\{1+\bs{1}_{\mathbb{B}(\bs{0},\sqrt{2B^{-1}})}(|\bs{x}-\bs{x}'|) \Big[ \Le|\ln
\Le(\tfrac{B}{2}|\bs{x}-\bs{x}'|^2\Ri)\Ri|  \nonumber\\
+\Le(1+\Le|\ln
\Le(\tfrac{B}{2}|\bs{x}-\bs{x}'|^2\Ri)\Ri|\Ri)|\bs{x}-\bs{x}'|^{-1}\Big]\Big\}
,\quad \alpha=1 \;  .
\end{cases}
\end{eqnarray}
Now, using $|\bs{x}-\bs{x}'|_\star \leq |\bs{x}-\bs{x}'|$, we can
replace the Euclidean distance with the distance $|\cdot|_\star$
in all the terms in the RHS of \eqref{ght}, since all these
functions are decreasing. To obtain the same bound for the terms
$|m|\leq 1$ in the sum we just drop the characteristic functions
$\bs{1}_{\mathbb{B}(\bs{0},\sqrt{2B^{-1}})}$.
\end{proof}

\section{Average velocity of the eigenstate associated to $E_\kappa^\alpha$}\label{appB}

In this appendix we prove following \cite{F} that the eigenstates corresponding to the
eigenvalues of $H_\alpha$ ($\alpha=\ell,r$) in a energy interval
$\Delta=(B-\delta,B+\delta)\subset
\Delta_\varepsilon$ have an average velocity that is strictly
positive/negative uniformly in $L$, that is, if we have
$H_\alpha\psi_\kappa^\alpha=E_\kappa^\alpha\psi_\kappa^\alpha$ then
\begin{equation}
|(\psi_\kappa^\alpha,v_y \psi_\kappa^\alpha)| \geq C'>0 \; .
\end{equation}

\noindent
>From the eigenvalue equation we have
\begin{equation}\label{un}
\|(H_\alpha^0-E^\alpha_\kappa)\psi_\kappa^\alpha\|^2
=\|V_\omega^\alpha\psi_\kappa^\alpha\|^2 \leq V_0^2 \; .
\end{equation}
We now expand $\psi_\kappa^\alpha$ on the eigenfunctions of
$H_\alpha^0$ denoted
$\Le\{\phi_{n,m}(x,y)=\frac{e^{iky}}{\sqrt{L}}\varphi_{nk}(x)\Ri\}_{n\in\N,k\in
\frac{2\pi}{L}\Z}$ where $\varphi_{nk}$ is the solution on the
eigenvalue problem
$[\frac{1}{2}p_x^2+\frac{1}{2}(k-Bx)^2+U_\alpha]\varphi_{nk}=E^\alpha_{nk}\varphi_{nk}$.
\begin{equation}\label{deuxtris}
\psi_\kappa^\alpha(x,y) = \sum_{n=0}^\infty \sum_{m\in \Z}
 \psi_n(m)\phi_{n,m}(x,y)\; ,
\end{equation}
and of course
\begin{equation}\label{trois}
\|\psi^\alpha_\kappa\|^2 = \sum_{n=0}^\infty \sum_{m\in\Z}
|\psi_n(m)|^2 = 1\; .
\end{equation}
>From \eqref{deuxtris} the equation \eqref{un} becomes
\begin{equation}\label{quatre}
\sum_{n=0}^\infty  \sum_{m\in \Z}|\psi_n(m)|^2
\Le(E_{n,m}^\alpha-E_\kappa^\alpha\Ri)^2   \leq V_0^2
\end{equation}
thus since each term in the sum is positive we have
\begin{equation}\label{cv}
\sum_{m\in \Z} |\psi_0(m)|^2 \Le(E_{0,m}^\alpha-E^\alpha_\kappa\Ri)^2
\leq V_0^2
\end{equation}
\noindent
We remark that for $n\geq 1$ one has $|E_{n,m}^\alpha-E_\kappa^\alpha|\geq
\frac{B}{2}-\delta$, this leads to
\begin{equation}\label{set}
\|\psi_\star\|^2\equiv \sum_{n=1}^\infty \sum_{m\in
\Z}|\psi_n(m)|^2 \leq \frac{V_0^2}{(\frac{B}{2}-\delta)^2} \; .
\end{equation}
Let $m^\star$ such that $|E_{0,m^\star}^\alpha-E_\kappa^\alpha|$ is minimal, and for a fixed $a$ independent of $L$ let
${\cal A}=[m^\star-a,m^\star+a]$. Then from \eqref{quatre}
\begin{eqnarray}
V_0^2 &\geq& \sum_{m\in
\Z} |\psi_0(m)|^2 \Le(E_{0,m}^\alpha-E_\kappa^\alpha\Ri)^2
\geq \sum_{m\in {\cal A}^c} |\psi_0(m)|^2 \Le(E_{0,m}^\alpha-E_\kappa^\alpha\Ri)^2 \nonumber \\
&\geq& \inf_{m\in {\cal A}^c} \Le(E_{0,m}^\alpha-E_\kappa^\alpha\Ri)^2
\sum_{m\in {\cal A}^c} |\psi_0(m)|^2
\end{eqnarray}
thus
\begin{equation}\label{supp omega}
\sum_{m\in {\cal A}^c} |\psi_0(m)|^2 \leq V_0^2\sup_{m\in{\cal A}^c}
\Le(E_{0,m}^\alpha-E_\kappa^\alpha\Ri)^{-2} \; .
\end{equation}
From \eqref{trois} and \eqref{set} we get
\begin{equation}\label{sito}
1 \geq \sum_{m\in \Z} |\psi_0(m)|^2 \geq 1
-\tfrac{V_0^2}{(\frac{B}{2}-\delta)^2}\; .
\end{equation}
Combining the last equation and \eqref{supp omega} we get
\begin{equation}
\sum_{m\in {\cal A}} |\psi_0(m)|^2
\ge 1-V_0^2 \Le[\tfrac{1}{(\frac{B}{2}-\delta)^2}+\sup_{m\in {\cal A}^c}(E_{0,m}^\alpha-E_\kappa^\alpha)^{-2}\Ri]
\; .
\end{equation}
Decompose now $\psi_\kappa^\alpha$ as $\psi_\kappa^\alpha=\psi_0+\psi_\star$, then
\begin{equation}
|(\psi^\alpha_\kappa,v_y\psi^\alpha_\kappa)| \geq |(\psi_0,v_y\psi_0)| -
|(\psi_\star,v_y\psi_\star)| - 2|(\psi_\star,v_y\psi_0)|
\end{equation}
the first term can be written as
\begin{eqnarray}
& &\int_\R \D x\, \int_{-\frac{L}{2}}^{\frac{L}{2}} \D y\,  \Le\{ \sum_{m'\in
\Z} \psi_0^*(m')\frac{e^{-i\frac{2\pi m'}{L}y}}{\sqrt{L}} \varphi_{0,m'}^*(x) \sum_{m\in
\Z} \psi_0(m) v_y \frac{e^{i\frac{2\pi m}{L}y}}{\sqrt{L}}\varphi_{0,m}(x) \Ri\}
\nonumber\\
&=& \sum_{m\in \Z}  |\psi_0(m)|^2  \int_\R \D x \Le(k-Bx\Ri)|\varphi_{0,m}(x)|^2
\nonumber\\
&=& \sum_{m\in \Z}  |\psi_0(m)|^2
\partial_{{k}}E_0^\alpha({k})\Big|_{{k}=\frac{2\pi m}{L}}
\end{eqnarray}
The partial derivative of $E_0^\alpha$ is the average velocity
$\partial_{{k}}E_0^\alpha({k})\Big|_{{k}=\frac{2\pi m}{L}}=J_{E^\alpha_{0,m}}$, thus
\begin{eqnarray}
|(\psi_0,v_y\psi_0)| &\geq& \Bigg|\sum_{m\in \Z}  |\psi_0(m)|^2
J_{E^\alpha_{0,m}}\Bigg| \nonumber \\
&\geq& |J_{E_{0\bar{m}}^\alpha}|\Le\{ 1
-V_0^2\Le[\tfrac{1}{(\frac{B}{2}-\delta)^2}+
\sup_{m\in {\cal A}^c}\Le(E_{0,m}^\alpha-E^\alpha_\kappa\Ri)^{-2}\Ri] \Ri\}
\end{eqnarray}
for a suitable $\bar{m}\in {\cal A}$, and we have
$|J_{E_{0,\bar{m}}^\alpha}|>0$.
The second term can be bounded as follows
$|(\psi_\star,v_y\psi_\star)| \leq
\|\psi_\star\|\|v_y\psi_\star\| \leq
\tfrac{V_0}{\frac{B}{2}-\delta}\|v_y\psi_\star\|$
and
\begin{eqnarray}
\|v_y\psi_\star\|^2&=&2\Le(\psi_\star,\tfrac{1}{2}\Le(p_y-Bx\Ri)^2\psi_\star\Ri)
\nonumber\\
&\leq&2\Le(\psi_\star,\Le[\tfrac{1}{2}p_x^2+\tfrac{1}{2}\Le(p_y-Bx\Ri)^2+U_\alpha\Ri]\psi_\star\Ri)\nonumber\\
&+&2\Le(\psi_0, \Le[\tfrac{1}{2}p_x^2+\tfrac{1}{2}\Le(p_y-Bx\Ri)^2+U_\alpha\Ri]\psi_0\Ri)
= 2\Le(\psi_\kappa^\alpha,H_\alpha^0\psi_\kappa^\alpha\Ri) \nonumber \\
&=&2(\psi_\kappa^\alpha,H_\alpha\psi_\kappa^\alpha)-2(\psi_\kappa^\alpha,V_\omega^\alpha\psi_\kappa^\alpha)
\leq 2(E_\kappa^\alpha+V_0) \; .
\end{eqnarray}
This leads to the bound
\begin{equation}
|(\psi_\star,v_y\psi_\star)| \leq
\tfrac{V_0}{\frac{B}{2}-\delta}\sqrt{2(E_\kappa^\alpha+V_0)}
\end{equation}
A similar argument gives the same bound for the third term.\\
Finally
\begin{eqnarray}
|(\psi_\kappa^\alpha,v_y\psi^\alpha_\kappa)| &\geq& |J_{E_{0,\bar{m}}^\alpha}|\Le\{ 1
-V_0^2\Le[\tfrac{1}{(\frac{B}{2}-\delta)^2} +
\sup_{m\in {\cal A}^c}\Le(E_{0,m}^\alpha-E^\alpha_\kappa\Ri)^{-2}\Ri] \Ri\} \nonumber \\
&-& 3
\tfrac{V_0}{\frac{B}{2}-\delta}\sqrt{2(E_\kappa^\alpha+V_0)}
\end{eqnarray}
that is strictly positive for a sufficiently small $V_0>0$ (we can remark that
the important condition is $V_0\ll B$).

\section{Discussion of hypothesis 1}\label{appC}

In this section we indicate a way in which hypothesis $(H1)$ can be achieved explicitly. We thank F. Bentosela for pointing out this
possibility to one of us. We take two
symmetric confining walls $U_\ell(-x)=U_r(x)\equiv U(x)$ and add 
 a magnetic flux tube of intensity $0\leq\Phi\leq 2\pi$  along the cylinder axis. Below we check that the magnetic flux
 lifts the degeneracy of the levels on the two sides of the sample. \\
In this case the pure edge Hamiltonians are
\begin{eqnarray}
H_\ell^0[\Phi]&=&\tfrac{1}{2}p_x^2 +
\tfrac{1}{2}\Le(p_y-Bx+\tfrac{\Phi}{L}\Ri)^2 +U(-x) \\
H_r^0[\Phi]&=&\tfrac{1}{2}p_x^2 +
\tfrac{1}{2}\Le(p_y-Bx+\tfrac{\Phi}{L}\Ri)^2 + U(x) \; .
\end{eqnarray}
The spectra of these Hamiltonians are
\begin{equation}
\sigma(H_\alpha^0[\Phi])=\{E^\alpha_{n,m}(\Phi): n\in \N , m\in
\Z \}.
\end{equation}
with $E^\alpha_{n,m}(\Phi)=\varepsilon_n^\alpha(\frac{2\pi m}{L}+\frac{\Phi}{L})$.
We consider here only the first spectral branches and note that 
from the symmetry of the walls, for $\Phi=0$
\begin{equation}
\varepsilon^\ell_0\Le(-\tfrac{2\pi}{L}m\Ri)=\varepsilon^r_0\Le(\tfrac{2\pi}{L}m\Ri) \quad
\forall \; m\in \Z
\end{equation}
We have
\begin{eqnarray}
\varepsilon^\ell_0\Le(-\tfrac{2\pi
m}{L}+\tfrac{\Phi}{L}\Ri)&=&\varepsilon^\ell_0\Le(-\tfrac{2\pi m}{L}\Ri) +
\partial_{k}\varepsilon^\ell_0(k_\ell)\frac{\Phi}{L} \\
\varepsilon^r_0\Le(\tfrac{2\pi m}{L}+\tfrac{\Phi}{L}\Ri)&=&\varepsilon^r_0\Le(\tfrac{2\pi m}{L}\Ri) +
\partial_{k}\varepsilon^r_0(k_r)\frac{\Phi}{L}
\end{eqnarray}
for a suitable
$\tfrac{2\pi}{L}(-m)\leq k_\ell\leq\tfrac{2\pi}{L}(-m)+\tfrac{\Phi}{L}$
and $\tfrac{2\pi}{L}m\leq k_r\leq\tfrac{2\pi}{L}m+\tfrac{\Phi}{L}$.
Thus
\begin{eqnarray}
\Le|\varepsilon^\ell_0\Le(-\tfrac{2\pi
m}{L}+\tfrac{\Phi}{L}\Ri)-\varepsilon^r_0\Le(\tfrac{2\pi m}{L}+\tfrac{\Phi}{L}\Ri)\Ri|
&=&\frac{\Phi}{L}\Le|\partial_{k}\varepsilon^r_0(k_r)-\partial_{k}
\varepsilon^\ell_0(k_\ell)\Ri| \nonumber\\
&\geq&2\frac{\Phi}{L} |\partial_{k}\varepsilon^\ell_0(k_\ell)| \geq 2{\cal C}\frac{\Phi}{L}
\end{eqnarray}
where ${\cal C}>0$.
A similar argument shows that
\begin{eqnarray}
& &\Le|\varepsilon^\ell_0\Le(-\tfrac{2\pi
(m+1)}{L}+\tfrac{\Phi}{L}\Ri)-\varepsilon^r_0\Le(\tfrac{2\pi
m}{L}+\tfrac{\Phi}{L}\Ri)\Ri| \nonumber \\
&=&\Le|\frac{\Phi}{L}\Le[\partial_{k}\varepsilon^\ell_0(k_\ell)-\partial_{k}
\varepsilon^r_0(k_r)\Ri]-\tfrac{2\pi}{L}\partial_{k}\varepsilon^\ell_0(k_\ell)\Ri|
\geq \Le|2\frac{\Phi}{L}
|\partial_{k}\varepsilon^\ell_0(k_\ell)|-\frac{2\pi}{L}|\partial_{k}\varepsilon^\ell_0(k_\ell)|\Ri|
\nonumber \\
&\geq& 2{\cal C}\frac{|\Phi-\pi|}{L}
\end{eqnarray}
Then, by fixing $\Phi^\star$ such that $0<\Phi^\star<\pi$ or
$\pi<\Phi^\star<2\pi$ we achive \eqref{hypot1}.

\section*{Acknowledgements} 
We wish to thank F. Bentosela, J.M. Combes, P. Exner, J. Fr\"ohlich and P.A. Martin for helpful discussions.
The work of C.F. was supported by a grant from the Fonds National Suisse
de la Recherche Scientifique No. 20 - 55654.98.


\begin{thebibliography}{2000}
\addcontentsline{toc}{section}{References}
\bibitem[BCD]{BCD} P. Briet, J.M. Combes, P. Duclos: Spectral stability under thunneling.
Commun. Math. Phys. {\bf 126}, 133 (1989)
\bibitem[BG]{BG} F. Bentosela, V. Grecchi: Stark Wannier Ladders. Commun.
Math. Phys. {\bf 142}, 169 (1991)
\bibitem[CH]{CH} J.M. Combes, P.D. Hislop: Landau Hamiltonians with random
potentials: localization and the density of states. Commun.
Math. Phys. {\bf 177}, 603 (1996)
\bibitem[CHS]{CHS} J.M. Combes, P.D. Hislop, E. Soccorsi: Edge states for
quantum Hall hamiltonians. Preprint mp-arc/02-172
\bibitem[dBP]{dBP} S. de Bi\`evre, J.V. Pul\'e: Propagating edge states for
magnetic Hamiltonian. Math. Phys. Electr. J. {\bf 5}, no. 3 (1999)
\bibitem[EJK]{EJK} P. Exner, A. Joye, H. Kovarik: Magnetic transport in a straight
parabolic channel. J. Phys. A: Math. Gen. {\bf 34}, 9733 (2001)
\bibitem[F]{F} C. Ferrari: Dynamique d'une particule quantique dans un
champ magn\'etique inhomog\`ene. Diploma work, EPFL (1999).
\bibitem[FGW]{FGW} J. Fr\" ohlich, G.M. Graf, J. Walcher:
On the extended nature of edge states of quantum Hall Hamiltonians.
Ann. Henri Poincar\'e {\bf 1}, 405 (2000)
\bibitem[FM1]{FM1} C. Ferrari, N. Macris: Intermixture of extended edge and
localized bulk energy levels in macroscopic Hall systems. J. Phys.
A: Math. Gen. {\bf 35} (scheduled August 2002)
\bibitem[FM2]{FM2} C. Ferrari, N. Macris: Spectral properties of
finite quantum Hall systems. To appear in the Proceedings of the Operator
Algebras and Mathematical Physics Conference (Constanta 2001, J.M.Combes,
J.Cuntz, G.E.Elliott, G.Nenciu, S.Stratila, H.Siedentop eds.),
published by the Theta Foundation. (Preprint mp-arc/02-121)
\bibitem[H]{H} B.I. Halperin: Quantized Hall conductance, current-carrying edge states,
and the existence of extended states in a two-dimensional disordered potential.
Phys. Rev. B {\bf{25}}, 2185 (1982)
\bibitem[M]{M} N. Macris: Spectral flow and level spacing of edge states for
quantum Hall Hamiltonians. Preprint math-ph/0206045
\bibitem[MMP]{MMP} N. Macris, P.A. Martin and J.V. Pul\'e: On Edge States
In Semi-Infinite Quantum Hall Systems. J. Phys. A: Math. Gen. {\bf{32}}, 1985 (1999)
\bibitem[PG]{PG} R.E. Prange and S.M. Girvin: \emph{The Quantum Hall Effect}.
New York: Graduate Texts in Contemporary Physics, Springer, 1987
\bibitem[vKDP]{vKDP} K. v. Klitzing, G. Dorda, M. Pepper: New method for high-accuracy determination of the
fine-structure constant based on quantized Hall resistance. Phys.
Rev. Lett. {\bf 45}, 494 (1980)
\end{thebibliography}
\end{document}